\documentclass[conference]{IEEEtran}
\IEEEoverridecommandlockouts

\usepackage{pbalance}

\IEEEoverridecommandlockouts 
\usepackage[dvips]{graphicx}
\usepackage{algorithmic}
\usepackage{algorithm}
\usepackage[cmex10]{amsmath}
\usepackage{url}
\usepackage{multirow}

\usepackage{amsmath}
\usepackage{amssymb}
\usepackage[T1]{fontenc}
\usepackage{graphicx}
\usepackage{float}
\usepackage{paralist}
\usepackage{xcolor}

\usepackage{pifont}

\begin{document}

\title{Is Homomorphic Encryption Feasible \\ for Smart Mobility?}
	
\author{\IEEEauthorblockN{Anika Hannemann}
\IEEEauthorblockA{Dept. of Computer Science, Leipzig University\\
Center for Scalable Data Analytics and Artificial\\
Intelligence (ScaDS.AI) Dresden/Leipzig, Germany \\
Email: anika.hannemann@informatik.uni-leipzig.de}
\and
\IEEEauthorblockN{Erik Buchmann}
\IEEEauthorblockA{
Dept. of Computer Science, Leipzig University\\
Center for Scalable Data Analytics and Artificial\\
Intelligence (ScaDS.AI) Dresden/Leipzig, Germany \\
Email: buchmann@informatik.uni-leipzig.de}
}	

%
%

\maketitle              
\begin{abstract}
Smart mobility is a promising approach to meet urban transport needs in an environmentally and and user-friendly way. Smart mobility computes itineraries with multiple means of transportation, e.g., trams, rental bikes or electric scooters, according to customer preferences. A mobility platform cares for reservations, connecting transports, invoicing and billing. This requires sharing sensible personal data with multiple parties, and puts data privacy at risk. 

In this paper, we investigate if fully homomorphic encryption (FHE) can be applied in practice to mitigate such privacy issues. FHE allows to calculate on encrypted data, without having to decrypt it first. We implemented three typical distributed computations in a smart mobility scenario with SEAL, a recent programming library for FHE. With this implementation, we have measured memory consumption and execution times for three variants of distributed transactions, that are representative for a wide range of smart mobility tasks. Our evaluation shows, that FHE is indeed applicable to smart mobility: With today's processing capabilities, state-of-the-art FHE increases a smart mobility transaction by about 100~milliseconds and less than 3~microcents. 
\end{abstract}



\section{Introduction}
\label{into}

Growing cities, urban sprawls and environmental concerns demand for mobility concepts~\cite{jevinger2019potentials,al2021enabling} that go beyond individual cars~\cite{schuppan2014urban}. Smart mobility is also known as multi-modal mobility or intelligent mobility. It refers to the integration of advanced technologies and intelligent systems into transportation networks to improve efficiency, safety, and sustainability, to lower emissions and to enhance the overall quality of urban life. To this end, smart mobility encompasses  solutions from the area of cloud computing, machine learning and artificial intelligence, that optimize the movement of people and goods within urban areas. 

With smart mobility, customers can specify a mobility demand and travel preferences~\cite{chowdhury2018preference}, e.g., the shortest, fastest, most inexpensive or eco-friendliest route from a starting point to a destination, or the route with the least transfers between the means of transportation. A cloud-based mobility platform~\cite{broring2017enabling} then lets the customer select an itinerary among some alternatives~\cite{li2017pare,herzog2017routeme,campigotto2016personalized,moran2017hybrid}.

For example, a customer might want to go from a train station to a stadium and must arrive at the beginning of a game. 
The mobility platform queries the databases of the connected mobility providers, and suggests three options that allow the customer to reach the destination in time: 
Based on an assessment of environmental impact, the most eco-friendly option is to use a tram to travel three stops to a rental bike station. As an alternative, a customer could opt to take a bus for six stops and use an electric scooter for the rest of the way. The most comfortable but expensive option would be hiring a cab. 
Once the customer has selected one option, the platform makes reservations. 
It collects the recorded distances and stations traveled. 
After the trip, the platform connects to services that handle invoicing and payment. 

Smart mobility approaches raise privacy concerns~\cite{paiva2020privacy,borchersprivacy,de2022impact}:
Start and end of a route can reveal personal needs, e.g., if it is a church, hospital or event location. 
It might be possible to identify an individual, by recurring ways home from work and vice versa. 
If customers frequently travel together, this indicates personal relationships. 
If a rental sports car is preferred over a suburb train, this might tell about preferences and wealth. 
Travel times and frequent routes reveal habits, employment status or daily routines. 
What makes privacy issues even more challenging is that personal data is distributed among many parties~\cite{eckhoff2017privacy}, such as
providers for mobility and infrastructure, and various services for payment, demand forecasting, parking space management, etc. 

Existing privacy-aware smart mobility approaches make use of anonymization, e.g. by using differential privacy~\cite{zhao2020survey,khaliq2022secure,qin2022toward} 
or by reducing the data resolution~\cite{shanthi2014efficient,memon2018pseudonym,martelli2022price}.
Alternatively, secure-multiparty computation can be used~\cite{li2019autompc,raja2020ai}. 
This induces noise to the data and might require multiple rounds of computation among several parties, i.e., it reduces accuracy, efficiency and, therefore, user experience. 

We investigate if fully homomorphic encryption~\cite{acar2018survey} (FHE) can be an alternative.
FHE allows to calculate sums or products on encrypted data without having to decrypt the data first. The decryption then provides the exact, noise-free result of the calculation. This makes FHE a natural choice for calculations with privacy-related data. However, some calculations increase the size of the data and/or make encryption, calculations or decryption computationally expensive. Some years ago, this limited the practical applicability of FHE in any real setting. 

However, vast advances in FHE programming libraries such as SEAL~\cite{sealcrypto23}, HElib~\cite{helib23} or OpenFHE~\cite{OpenFHE23}, new paradigms such as edge computing, and a huge computing power available at little costs both at edge nodes and cloud services 
give reasons for new analyses. For example, in May 2023 a cloud instance with 128 XEON CPU cores at 3.5 GHz and 512 GiB RAM and 50 Gbps network bandwith costs only 8~USD per hour. Thus, our concern is to find out whether the overhead of FHE for typical smart mobility transactions is reasonable to support privacy-compliant business models in this appliaction domain.

In this paper, we make the following contributions: 

\begin{itemize}
\item We identify three distributed transactions that are representative for a smart mobility scenario and benefit from FHE, i.e., require noise-free results and cannot be readily secured by simpler means such as one-time pseudonyms~\cite{memon2018pseudonym}. 
\item We implement a prototype based on Microsoft SEAL~\cite{sealcrypto23}, which uses the state-of-the-art FHE schemes BGV~\cite{cryptoeprint:2012/144}, BFV~\cite{cryptoeprint:2011/277} and CKKS~\cite{cheon2017homomorphic}.
\item We measure memory consumption and execution times, and we compare them with the resources available in the cloud or on a smartphone. 
\end{itemize}

Our evaluation shows that with CKKS, encrypted transactions add approx. 100~ms to the CPU time of unencrypted transactions. This does not impact user experience~\cite{nielsen2013mobile}. With parallel processing, this time can be reduced, and it costs less than 3~microcents on a current cloud instance. Thus, we have confirmed that FHE is indeed feasible for smart mobility  business models, where such transaction fees are several orders of magnitude smaller than the billing amount on the customer’s invoice, but privacy is an important factor.

\noindent\textbf{Paper structure}: 
The next section reviews related work. 
In Section~\ref{sec:approach}, we derive our smart mobility transactions with fully homomorphic encryption. 
Section~\ref{sec:eval} contains an experimental evaluation, and Section~\ref{sec:conclusion} concludes. 


\section{Related Work}
\label{sec:rel}

In this section, we explain the state-of-the-art in (fully) homomorphic encryption, and we briefly review smart mobility approaches. 

\subsection{Homomorphic Encryption}
Homomorphic Encryption (HE) is a well-established technique that enables third-party computation on encrypted data, without requiring the data to be decrypted beforehand. HE allows for data to be encrypted while keeping the features of the function and format of the encrypted data, supporting  privacy-preserving data processing. Although this property of HE is known already for over than 30 years, 
the first plausible FHE approach was proposed by Gentry et al. in 2009 ~\cite{gentry2009fully}. However, HE is costly in terms of computation and is, therefore, still subject of ongoing research~\cite{acar2018survey}. 

The homomorphic property of HE allows certain operations to be computed over the encrypted data, with the resulting values also being encrypted. For two messages $\forall m_1, m_2 \in \mathcal{M}$ of a message space $\mathcal{M}$, Eq. \ref{HE} shows an HE scheme that supports any operation on their respective ciphers $c_1, c_2$. In the context of a public-key cryptosystem, the public key is denoted as $k_e$, the private key as $k_d$, and the encryption and decryption functions as $Enc$ and $Dec$, respectively. 

\begin{equation}
\begin{aligned}\label{HE}
    c_1=Enc(k_e, m_1), \ \ c_2 = Enc(k_e, m_2) \\
    m_1 \oplus m_2 = Dec(k_d, c_1 \oplus c_2)
\end{aligned}    
\end{equation} 

Formally, a homomorphic encryption scheme is a quadrupel  $HE$=($KeyGen$, $Enc$, $Dec$, $Eval$) with $KeyGen$, $Enc$, $Dec$ and $Eval$ being probabilistic polynomial time algorithms: \\
\begin{compactitem}
    \item[\textit{\textbf{KeyGen}}] generates a public key $k_e$, a private key $k_d$ and an evaluation key $k_{eval}$ given some security parameter $\lambda$ for the asymmetric version of HE: $KeyGen(1^\lambda) \rightarrow (k_e, k_d, k_{eval})$ \\
    For the symmetric version, only a secret key $k_d$ and an evaluation key $k_{eval}$ are created. 
    \item[\textit{\textbf{Enc}}] encrypts a plaintext message $m \in \mathbb{Z}_n$ to a ciphertext $c$ using the public key, which is shared:  $Enc(k_e, m) \rightarrow c$ 
    \item[\textit{\textbf{Dec}}] uses the private key, which is kept secret, to decrypt a ciphertext $c$ to a plaintext message $m$: $Dec(k_d, c) \rightarrow m$ 
    \item[\textit{\textbf{Eval}}] applies an operation $f:\mathbb{Z}_n^l$ $\rightarrow$ $\mathbb{Z}_n$ to a given ciphertext $c_1, ..., c_l$ and outputs a ciphertext $c_f$ using the evaluation key $k_{eval}$: 
    $Eval(k_{eval}, f, c_1, \cdots, c_l) \rightarrow c_f$ \\
    $k_{eval}$ is generated uniquely for every computation and, therefore, does not pose a privacy threat. With $Eval$, homomorphism of the scheme can be proven.
\end{compactitem}

\subsection{Fully Homomorphic Encryption}

Depending on the support of the operations applied in the $Eval$ function, HE can be categorized into 
\textbf{fully homomorphic encryption} (FHE), \textbf{partially homomorphic encryption} (PHE) and \textbf{somewhat homomorphic encryption} (SHE), each of them with different limitations and capabilities. PHE allows $Eval$ for one operation $\oplus$, either addition of multiplication, for an unlimited number of times. SHE allows both addition and multiplication, but with a limited number of operations due to the increasing size of the ciphertext. FHE allows an unlimited number of operations $\oplus$ for an unlimited number of times. Addition and multiplication operations as well as comparison and branching are supported. Therefore, FHE is the most powerful approach of HE and, therefore, implemented in this work. 

FHE is a form of ring homomorphism with structure preserving characteristics~\cite{gentry2009fully}. This allows for arbitrary computations to be performed, as the homomorphic properties of the ring ensure that the results of the computations can be obtained without requiring decryption.  
Which operation is allowed depends on the FHE scheme; in this work the well known Brakerski-Fan-Vercauteren (BFV)~\cite{cryptoeprint:2011/277}, Brakerski-Gentry-Vaikuntanathan  (BGV)~\cite{cryptoeprint:2012/144} and Cheon-Kim-Kim-Song (CKKS)~\cite{cheon2017homomorphic} schemes are implemented and evaluated. They are based on on the hardness of the (Ring) Learning With Errors (RLWE) problem. Learning with Errors is considered to be one of the hardest, post-quantum problems to solve in polynomial time: Given $(x, y)$ where $y=f(x)$ for some linear function $f$, $f$ can be easily learned. Now, when adding errors to the algorithm's input such that $y \neq f(x)$ for a small probability, it is assumed that the problem can not be solved in polynomial time and is, therefore, hard~\cite{regev2006lattice}. 

A subproblem is the Ring Learning with Errors (RLWE), an extension of the LWE problem for polynomial rings over finite fields. A major advantage of RLWE is the key size: While the private and public keys of LWE-based cryptography can become large, RLWE-based keys are roughly the square root of LWE~\cite{lyubashevsky2013ideal}.

For FHE, there are methods for maintaining the ciphertext, without modifying the message, such as bootstrapping and relinearization. In bootstrapping, the evaluation key $k_{eval}$ is used to control noise. Thus, any number of FHE operations can now be computed without noise becoming uncontrollable. In this context, noise refers to a measure to prevent unauthorized decryption of encrypted data using the secret key, and it does not affect the precision of the computation outcome.
Relinearization handles a common problem of RLWE-based FHE, whose ciphertext sizes increases with every
homomorphic multiplication. During homomorphic evaluation, relinearization limits the expansion of the ciphertext to prevent high computation costs. 

The variety of supported operations allow for a wide range of computations to be performed on encrypted data, making FHE powerful and versatile and applicable in multiple settings. In Cloud Computing, FHE is used to protect the client's data privacy to process them on an external party~\cite{behera2022design,gupta2022memfhe}. The line of FHE works on Machine Learning aim to protect the training data's privacy in either a collaborative setting~\cite{lee2022privacy,chen2021privacy} or a federated learning setting~\cite{wibawa2022homomorphic,stripelis2021secure}. Another relevant application specifically for this work is private fog computing for the Internet-of-Things (IoT) which enables multiple users to authenticate and aggregate data collected with edge devices~\cite{zhu2019privacy}. 

\subsection{Smart Mobility}

The use of IoT technologies has proven to be an appropriate response to the growth of cities and the associated impact on traffic and transportation. It has brought up the concept of smart mobility, which refers to the optimal combination of various modes of transportation, including e-bikes, e-rollers, buses, shared cars, tramways, and trains, as well as infrastructure components such as roads, bridges, airports, and train stations. As transportation modes continue to grow and become more interconnected, the resulting complexity can make it increasingly challenging to efficiently use and combine available options. To tackle this problem, relevant related work has been done concerning urban mobility and multi-modal routing planning.~\cite{li2017pare,herzog2017routeme,campigotto2016personalized,moran2017hybrid} proposed a mobile recommender system for personalized multi-modal routes by utilizing a hybrid approach combining various IoT devices, primarily targeted for private cabs and taxis. The focus of a contribution of Al-Rahamneh et a. is on creating an multi-modal urban data platform with context-awareness~\cite{al2021enabling}. 

\cite{jevinger2019potentials} provides an analysis on the potentials of multi-modal travel support, but does not provide a framework or architecture. A mathematical model for preference-aware transport matching is contributed by~\cite{chowdhury2018preference}. The European Platforms Initiative project BIG IoT has been initiated to implement smart mobility services and applications for Barcelona, Piedmont, and Berlin/Wolfsburg. It aims to solve the interoperability gap by defining a generic, unified Web API for smart object platforms~\cite{broring2017enabling}. 
Smart mobility is of interest to both the research community and society. Nevertheless, there are are only a few studies targeting smart mobility with Homomorphic Encryption. \cite{ren2021privacy,baharon2015new} provide a privacy preserving solution for mobile cloud computing using IoT devices with HE. However, they do not propose a framework or analysis of the application on smart mobility. 


\section{Smart Mobility with Fully Homomorphic Encryption}
\label{sec:approach}

In this section, we introduce our system architecture, derive privacy requirements, and describe transactions that are representative for smart mobility and can be implemented with FHE. 

\subsection{Smart Mobility Architecture}
\begin{figure*}[ht]
\centering
	\includegraphics[width=1.5\columnwidth]{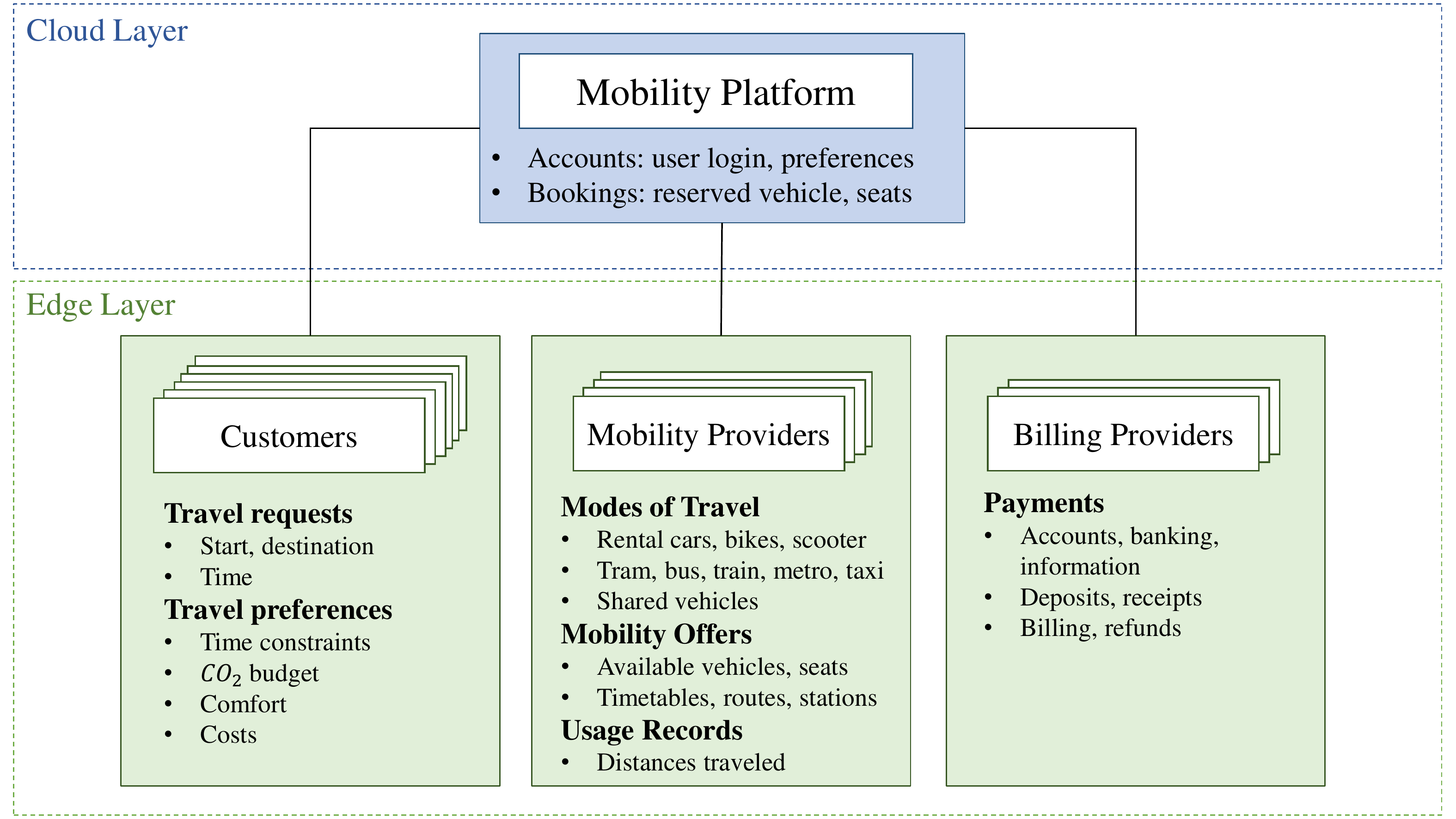}
	\caption{Smart Mobility Architecture}
	\label{img:smartmob}
\end{figure*}

To find out if FHE is applicable to smart mobility, we use a generalized architecture model, as shown in Figure~\ref{img:smartmob}. This model is in line with existing work~\cite{broring2017enabling}.

\textbf{Customers} issue travel requests, that have a start point, a destination and a start/end time. The travel requests can be constrained by the customer's budget and preferences regarding speed, comfort, eco-friendliness, etc. 

The \textbf{mobility platform} manages user accounts, connects all parties with each other, and provides platform services such as identifying potential mobility providers for a travel request and booking seats or vehicles. 

\textbf{Mobility providers} deploy various means of transportation, e.g., rental/shared vehicles or seats in a public transportation service. Furthermore,  mobility providers log the actual distances traveled with each vehicle or on each seat. 

Finally, \textbf{billing providers} invoice the trips made by each means of transportation. 

Observe, that in this architecture only the customers are natural persons in the sense of the GDPR~\cite{gdpr}. All other parties are institutions, that are not covered by the data protection regulations. Thus, only data related to customers must be protected.

\subsection{Privacy requirements}
The components of our architecture model process five distinct categories of data: 

\textbf{Insensitive data} cannot be related to a person, and does not carry sensible information. Any information from an institutional party such as the mobility platform or a mobility provider is insensitive data, e.g, the public keys of those parties. We also consider the aggregated values calculated with FHE to be insensitive, e.g., total travel costs, duration of a trip or CO$_2$ budget. 

\textbf{Identifiers} such as a name or a bank account reveal the identity of a person. A trip can be an identifier, if ends at the customer's home. 

\textbf{Pseudonyms} such as a login name can be changed easily. We assume that a pseudonym only allows to recognize a person during one transaction. The public key of a customer is also a pseudonym. 

\textbf{Sensitive data}, refers to personal information, e.g., habits, social life (persons traveling together), mobility preferences or travel costs. Note that sensitive data is not necessarily identifying. 

Finally, \textbf{secret data} includes all information that must not be shared, e.g., the private keys of the various parties. 
From this categorization, we derive three privacy requirements, as summarized in Table~\ref{tab:privreq}: 

\begin{itemize}
\item[\textbf{R1}] The mobility platform connects customers with two kinds of providers. Therefore, it needs to maintain user pseudonyms and transaction IDs. The platform does not need to learn identifiers or sensitive data, e.g., travel data forwarded to mobility providers. 
\item[\textbf{R2}] 
Mobility providers must learn which vehicles or seats are booked for which periods of time, and where a rented vehicle is left at the end of the trip. To create an invoice, the actual usage must be recorded. This requires pseudonymous information and sensitive data. 
It must be impossible to join sensitive data from multiple transactions or across multiple mobility providers.
\item[\textbf{R3}] A billing provider needs to know identifiers (names, addresses, bank accounts) and invoice amounts. 
It is also acceptable if the billing provider learns pseudonyms. Except from that, it should learn only insensitive data. 
\end{itemize}

We assume that a combination of one-time pseudonyms and traditional encryption 
helps to mitigate any privacy problem that relates to data-management transactions, e.g., marking a certain seat in a database as ``reserved'', searching for available modes of transportation at the last stop of a tram, or recording the time a rental bike has been used. 

This leaves open privacy issues related to calculations. For example, consider the billing process. In traditional smart mobility scenarios, the cloud platform might calculate the invoice total by asking each mobility provider, that was involved in the trip of a certain customer. By doing so, the smart-mobility platform learns the exact movement patterns of each customer. 

FHE might be able to execute such calculations without revealing personal details. The advantage of using FHE over alternatives from the realm of secure multiparty computation is, that FHE does not depend on privacy models where multiple semi-honest parties execute protocols in multiple rounds of communication. If each transaction is secured with its own pair of one-time keys, the security and privacy of the approach only depend on the formal guarantees of the FHE schemes used. We also do not need to make assumptions about colluding parties.\\[0.5cm]

\begin{table}[ht]
	\centering
		\begin{tabular}[h]{|l|c|c|c|}\hline
                    & \textsl{Mobility} & \textsl{Mobility} & \textsl{Billing}  \\
                    & \textsl{Platform} & \textsl{Provider} & \textsl{Provider}  \\                    
\textsl{Data Categories} &  (R1) &  (R2) & (R3) \\\hline
Insensitive Data    & \ding{51} & \ding{51} & \ding{51} \\ \hline
Identifiers         & \ding{55} & \ding{55} & \ding{51} \\ \hline
Pseudonyms          & \ding{51} & \ding{51} & \ding{51} \\ \hline
Sensitive Data      & \ding{55} & \ding{51} & \ding{51} \\ \hline
Secret Data         & \ding{55} & \ding{55} & \ding{55} \\ \hline
		\end{tabular}
  \vspace{15pt}
    \caption{\label{tab:privreq} Privacy Requirements}
\end{table}

\subsection{Experiment Design}
As we explore the applicability of FHE for smart mobility, we rule out transactions without sensitive/pseudonymous data or where encryption, decryption and computations take place on a cloud instance. An application of the privacy requirements (Table~\ref{tab:privreq}) on our architecture model (Figure~\ref{img:smartmob}) has shown that the billing provider has similar properties as the mobility provider: It learns identifiers instead of pseudonyms, and it has comparable computational means and data flows. Thus, we also leave aside experiments that specifically address billing providers. We experiment with three transactions T1--T3. Each transaction contains a small number of additions and multiplications. This is typical for business transactions that compute arrival times, discounts or usage fees. With FHE, such a transactions require one relinearization operation.

\begin{figure}[ht]
\centering
	\includegraphics[width=\columnwidth]{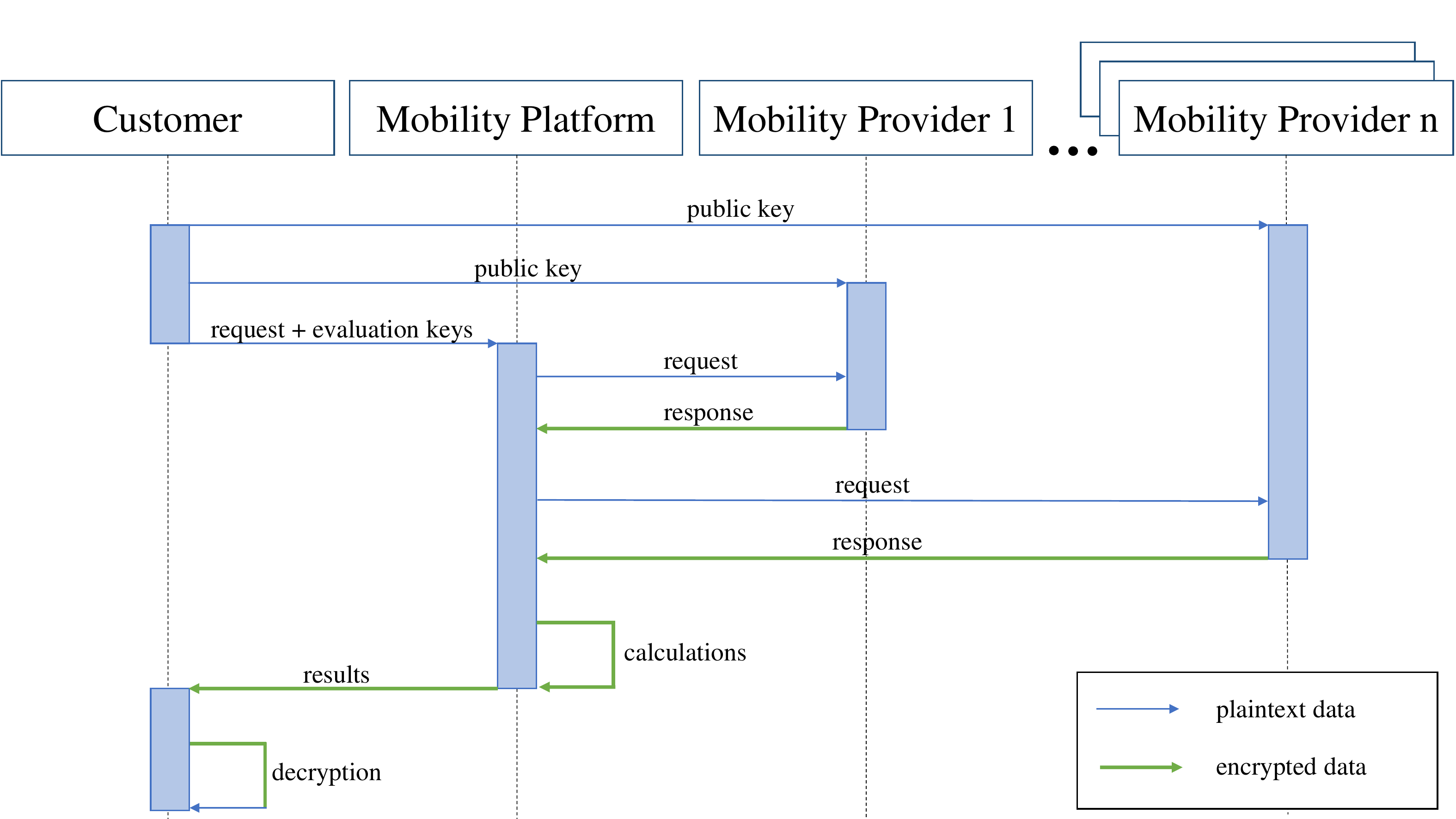}
	\caption{T1: Centralized Calculations}
	\label{img:t1}
\end{figure}

\paragraph{T1: Centralized calculations.} This transaction is representative for operations where encryption and decryption takes place at the customer's smartphone, while the calculation is executed at the mobility platform. For example, the mobility platform might add up encrypted prices, travel times, CO$_2$ budgets, etc., whose summands stem from the mobility providers. It might also multiply discounts or subtract bonuses. Then the mobility platform sends the encrypted result to the customer for decryption. 
The mobility providers might not want to reveal bonus schemes, mutual price agreements or internal calculations. Thus, it is not an option to send plain values to the customer and let the smartphone do any calculation. Instead, FHE can be applied. 
Figure~\ref{img:t1} illustrates T1. Thick lines in the figure refer to encrypted data. 

With this transaction, numerous parallel transactions must not overload the computational resources of the mobility platform, and the encryption/decryption of a single transaction must be feasible on the customer's smartphone. 
T1 requires the mobility providers to know the public key of the customer. This can be an one-time key, and it is a pseudonym. 
Each mobility provider only learns which of its own means of transport is part of the transaction. Thus, R2 is met. 
The mobility platform does not learn the customer's the public key, because relinearization requires evaluation keys that are used only once. 
To manage travels across multiple mobility providers, the mobility platform needs a transaction id, that is an one-time pseudonym. Thus, R1 is fulfilled, too.

\begin{figure}[ht]
\centering
	\includegraphics[width=\columnwidth]{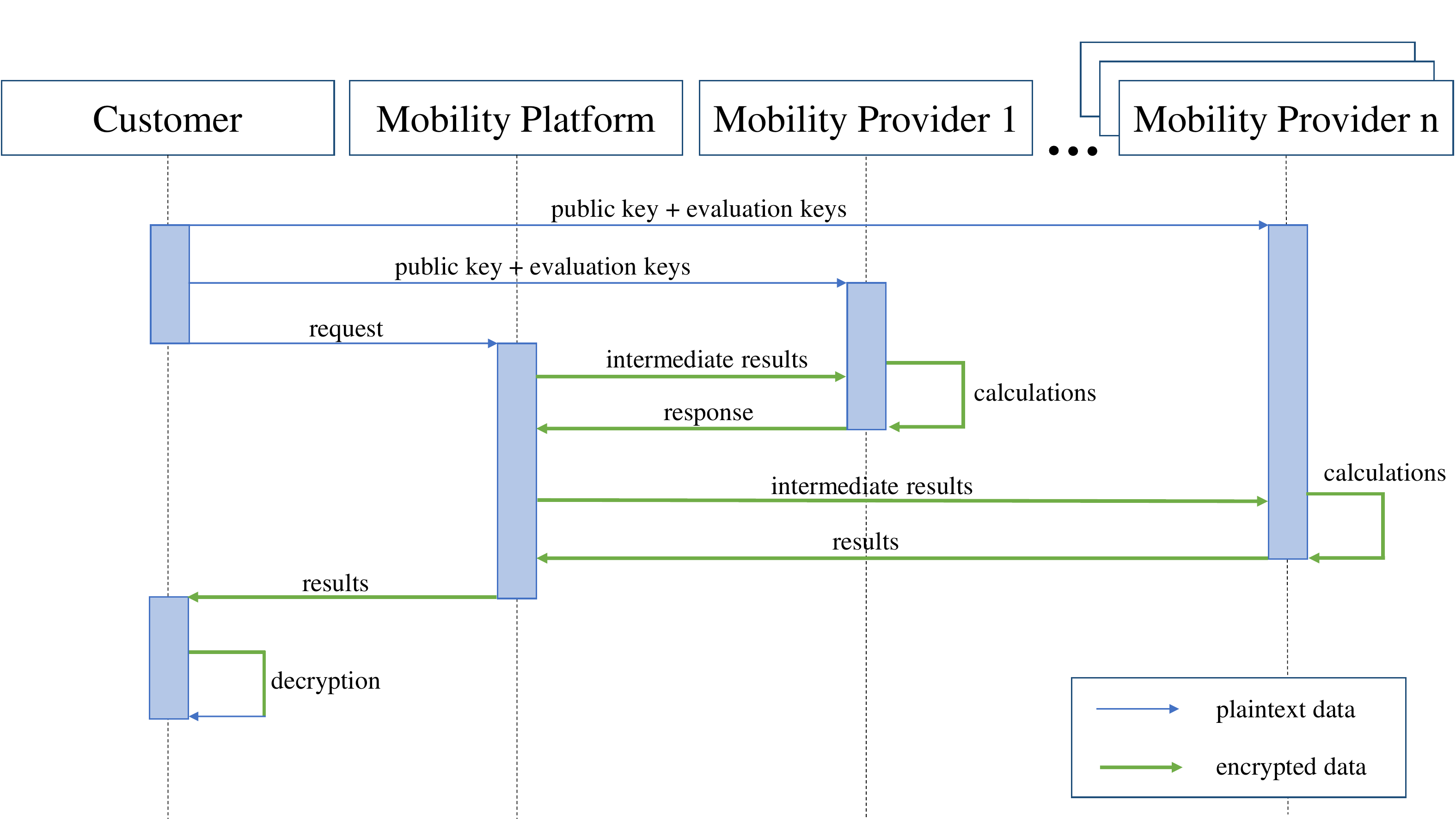}
	\caption{T2: Decentralized Calculations}
	\label{img:t2}
\end{figure}

\paragraph{T2: Decentralized calculations.} This transaction lets the mobility providers do the calculations (cf.~Figure~\ref{img:t2}). The mobility platform transfers encrypted intermediate results to different parties, e.g., to a rental car provider or a public transport provider, and orchestrates distributed computations on encrypted data there. Such parties have smaller computational resources than the mobility platform, but are also loaded with a smaller number of parallel transactions. 
Encryption/decryption takes place at the customer's smartphone. 

Similarly to T1, the reason for using FHE is, that the mobility providers might not want to share internal agreements and calculations. 
Again, the mobility providers need the customer's public key. Thus, the privacy properties of T2 are identical to T1, but the resources needed at the various parties are different. 

\begin{figure}[ht]
\centering
	\includegraphics[width=\columnwidth]{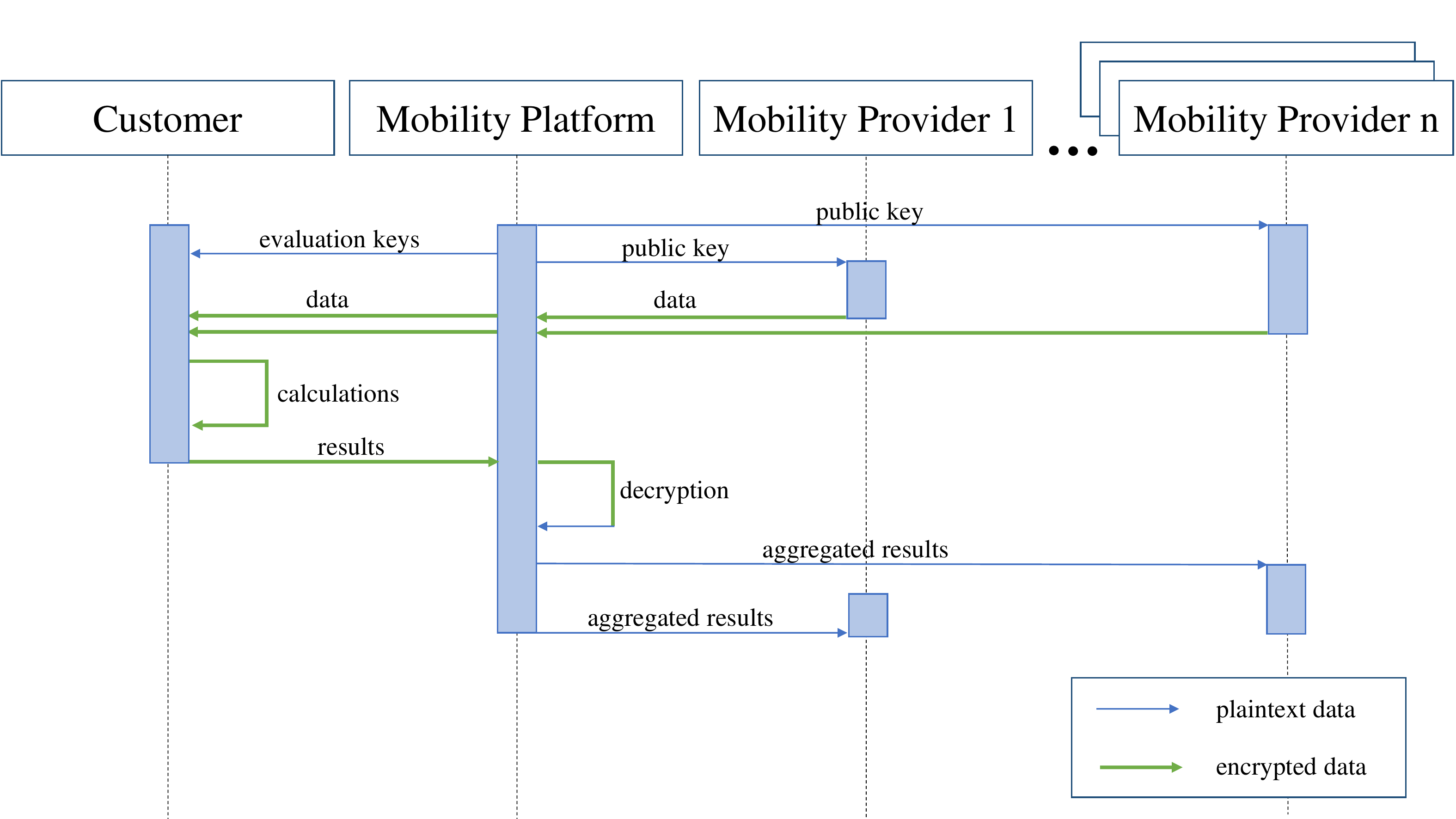}
	\caption{T3: Customer-side Calculations}
	\label{img:t3}
\end{figure}

\paragraph{T3: Customer-side calculations.} Our third transaction performs calculations at the customer's smartphone, while the mobility providers contribute encrypted parameters, and the  mobility platform is responsible for decryption (cf.~Figure~\ref{img:t3}). 
Thus, the reason for requiring FHE is similar to T1 and T2.
Such a transaction could calculate with telemetry data: Forecasting demand, costs, CO$_2$ per hour etc. require usage-dependent calculations with data from multiple mobility providers. But it might be sufficient for each mobility provider to learn the aggregated numbers. 

With T3, the calculations must not overload a smartphone, and decrypting many results in parallel must be feasible for a mobility platform. 
T3 means that the mobility providers need the public key of the mobility platform. Because it is an institutional party, this is insensitive data (cf.~Table~\ref{tab:privreq}). The mobility platform cannot learn which parameters from which mobility provider contribute to the aggregated results. Similarly, mobility providers only learn aggregated results, which is insensitive data. Thus, privacy requirements R1 and R2 are fulfilled.  

Our three transactions are representative for a wide number of typical real-world calculations in smart mobility scenarios. 
Therefore, we refrain from realizing other transactions that have the same structure and do not provide further insights. 
For example, in the billing process the invoice amounts could be encrypted by the mobility providers, calculated at the platform and decrypted by a billing provider, which has the same structure as T1. 


\section{Experimental Evaluation}
\label{sec:eval}

In this section, we define the computational and financial overhead we deem acceptable for smart mobility. 
Furthermore, we describe our prototypical implementation of our transaction, and we evaluate it with a series of experiments. 

\subsection{Resources and Costs}
To find out, if FHE can be applied to smart-mobility scenarios, we need an understanding of the resources available and the costs involved.
The mobility platform is a cloud service with plenty of computational resources for massive parallel processing. However, it must handle a very large number of requests at the same time. Furthermore, some FHE operations cannot be parallelized. 
Mobility providers are part of the edge layer of our architecture model. Thus, such providers would operate its services on a cloud instance that is one order of magnitude smaller and less expensive than those of the mobility platform. As multiple mobility providers exist, each of them has a much smaller individual load of parallel transactions. 
%
Customers connect via smartphone to the mobility platform. A smartphone has comparatively scarce computing resources. However, as it is the customer's property, it does not need to execute multiple transactions in parallel. 

As a reference for computing times on a current mid-range smartphone, consider a ``Fairphone 4''~\cite{fairphone23}. It has a CPU with 8 cores at 2.2~GHz and 8~GiB RAM and 128~GiB internal storage. An Amazon AWS instance ``m6i.32xlarge''~\cite{amazonaws23} serves as a reference for the computing costs of a large cloud instance. It is equipped with 128 XEON CPU cores at 3.5 GHz, 512~GiB RAM, 50~Gbps network bandwith. In January 2023, a m6i.32xlarge instance costs approx. 8 USD per hour, i.e., one second of one core costs 0,017~millicents. 

We assume that it is acceptable for a customer as well as for any other party, if a transaction takes at most two seconds to complete. 
This time is comparable to starting an app on a smartphone, i.e., it does not impact the user experience. 
For comparison~\cite{nielsen2013mobile}, humans do not perceive a reaction time below 400~ms as an interruption, and cannot sense delays below 100~ms. Furthermore, a few seconds computing time on a small number of cores of a large cloud instance does not contribute much to the total travel costs of the customer. 


\subsection{Implementation}
For each of our three distributed transactions, we have implemented a FHE-encrypted variant, and a non-encrypted one for comparison. 
We decided to implement our transactions in C++ with Microsoft SEAL~\cite{sealcrypto23}, because it is the most advanced implementation of the three state-of-the-art schemes BGV~\cite{cryptoeprint:2012/144}, BFV~\cite{cryptoeprint:2011/277} and CKKS~\cite{cheon2017homomorphic}. While the first two schemes compute with integers, the last one supports float-point operations. The length of the integers and the precision of float-point operations depend on the size of the modulus degree. The modulus determines how much noise can be accumulated during computations, before a relinearization operation with a pre-calculated evaluation key is needed. The noise is an internal measure to avoid that encrypted data can be decoded without knowing the secret key, i.e., it has nothing to do with the accuracy of the computation result. We used a polynomial modulus degree of 16384 and a plain modulus degree of 1024, which is the recommended setting in SEAL. For performance reasons, we disabled the debug mode and enabled batch processing, i.e., the SEAL library did not encrypt or decrypt any value individually. 

We executed our experiments on a host with a 2.8 GHz Intel i7 CPU with 8 cores and 32~GiB RAM. Thus, one CPU core of our experimental host is approx. 25\% slower of a core of an ``m6i.32xlarge'' instance and 30\% faster than one core of a ``Fairphone 4''. We have started one customer, one mobility platform and two mobility providers as separate processes, and we have repeated each experiment 500 times and computed the averages. 

We want to measure the processing time and the size of the encrypted data at each party separately. This allows us to find out if a FHE scheme exceeds the time budget of 2-3 seconds in total, or if one of the parties might be potentially drained of resources, when handling many customers in parallel. 

In order to execute the measurements, we have implemented our experiments as testcases with the DOCtest framework~\cite{doctest22}. This allows to implement experiments as a batch, and to verify that the computed results are correct. We used log4cplus~\cite{log4cplus22} to monitor the execution, and we measured with the Google Benchmark v1.7.1~\cite{googlebench22} microbenchmarking framework. Google Benchmark ensures that the compiler does not change the execution, e.g., to optimize 500 repetitions of the same execution. It delivers the values measured in a JSON format. We also evaluated the individual method calls with Intels VTUne profiler~\cite{vtune}. 

\subsection{Evaluation Results}
First we analyze the performance of BGV, BFV and CKKS. For comparability, we have measured the runtimes as CPU time on a single core. After that, we measure the memory consumptions. 

\begin{figure*}[ht]
\centering
\begin{minipage}[t]{0.3\columnwidth}
  \raggedright
	\includegraphics[height=6.5cm]{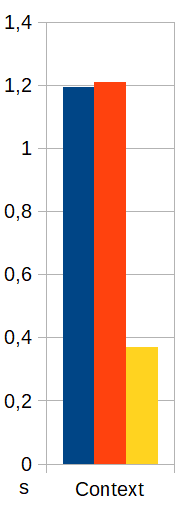}
	\caption{Runtimes of the Context Creation}
	\label{img:op-context}
\end{minipage}%
\begin{minipage}[t]{0.7\columnwidth}
  \raggedleft
	\includegraphics[height=6.5cm]{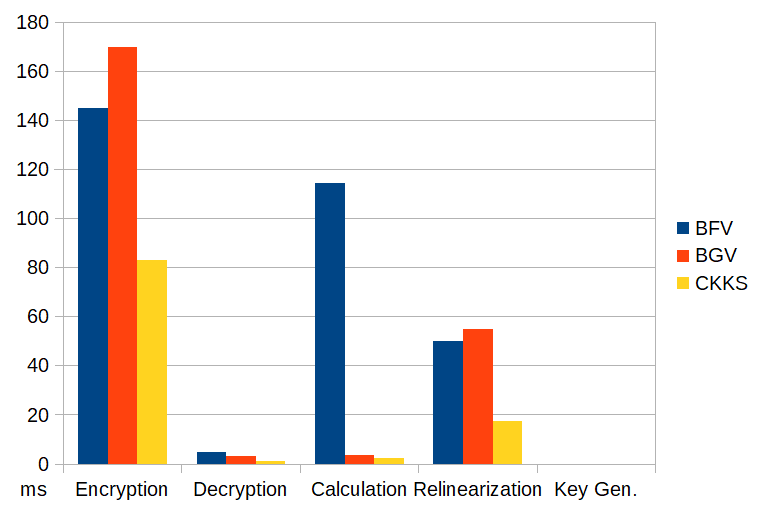}
	\caption{Runtimes of the Operations}
	\label{img:operations}
\end{minipage}
\end{figure*}

\paragraph{Runtime Performance}
Context creation takes only once at startup of a service or application. The purpose of this operation is to initialize and configure the SEAL library with the appropriate credentials, seeds, buffers etc. for the respective FHE scheme. As Figure~\ref{img:op-context} shows, we have measured an average context-creation time of up to 1.2~s. Thus, it is mandatory for any application, not to shut down and start up the FHE library for each operation, but to preserve its state. Note that context creation can be executed in parallel to the normal start-up of an application. Since even today's smartphones have multiple CPU cores, this overhead does not necessarily increase the application's startup  time. 


Figure~\ref{img:operations} shows the runtimes in milliseconds of the FHE operations. 
Encryption and decryption refer to the respective cryptographic operations. 
With calculation, we denote to a basic mathematical operation consisting of a few additions, multiplications and subtractions. Surprisingly, this was a time-consuming operation with BFV, which took approx. 115~ms on our 2.8~GHz Intel i7 CPU. The other schemes required 4~ms and 2.3~ms. 
Relinearization is needed after some calculations, to ensure that the decryption produces correct results. The relinearization requires an one-time evaluation key, whose creation time is depicted in the last column of Figure~\ref{img:operations}. 

As the figure shows, CKKS consumes approximately half the CPU time of the two other schemes, and the most expensive operations are encryption and relinearization. The runtimes for computations on encrypted data are orders of magnitude higher than on plain-text values. However, humans do not perceive reaction times below 400~ms as annoying, and do not recognize a delay below 100~ms at all~\cite{nielsen2013mobile}. Note that our measures only consider the runtimes of the FHE operations, i.e., we leave aside context creation, operating system, start-up times of applications and delays of network connections. 

We want to find out if those runtimes add up to a disruptive amount for our three transactions T1, T2 and T3. Therefore, we measured and aggregated the CPU times for any operation on any party, again for each of the schemes BFV, BGV and CKKS. We have left aside the context creation. 

To avoid confounding the effects of parallelization with the CPU times needed at the various parties, we have structured our experiment so that all mobility providers operate independently in parallel, while the customer and the mobility platform wait for all other parties. For the same reason, we do not measure network delays, effects of the operation system, etc. In a real setting, each party would start encrypting, decrypting or calculating values as soon as the first chunk of data has arrived, i.e., the total runtimes would be smaller. 

To foster comparability among the transactions, we also ensured that T1-T3 used the same set of operations, and differed only in the place where each encryption, decryption, calculation etc. took place. 
The set of operations contained a number of additions and calculations that was large enough to require one relinearization. 
Having said this, the BFV scheme required a total of 314~ms CPU time on average to complete a transaction. BGV needed an average of 232~ms, and CKKS had the best runtime performance with only 104~ms on average. 


\begin{figure}[ht]
  \centering
	\includegraphics[height=6cm]{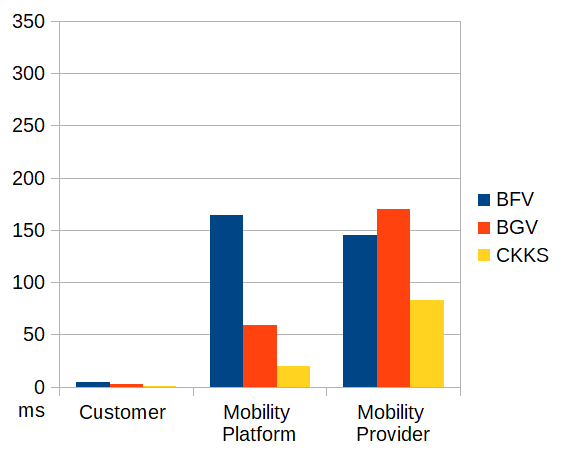}
	\caption{T1: Centralized Calculations}
	\label{img:t1calc}
\end{figure}

\begin{figure}[ht]
 \centering
	\includegraphics[height=6cm]{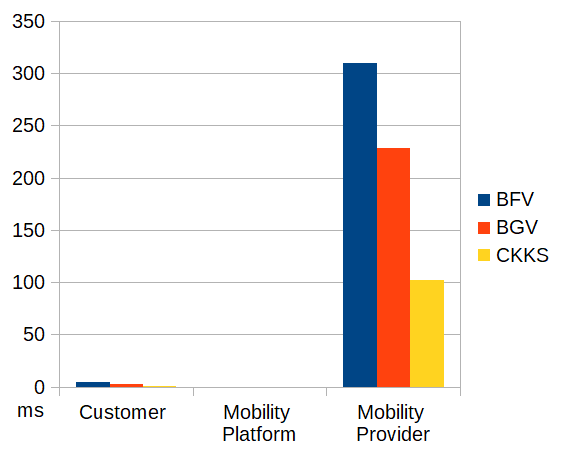}
	\caption{T2: Decentralized Calculations}
	\label{img:t2calc}
\end{figure}

\begin{figure}[ht]
  \centering
	\includegraphics[height=6cm]{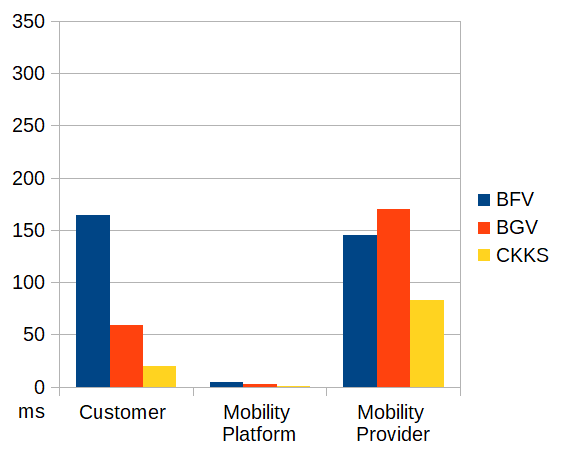}
	\caption{T3: Customer-side Calculations}
	\label{img:t3calc}
\end{figure}

Figures~\ref{img:t1calc}-\ref{img:t3calc} show for centralized, decentralized and customer-side calculations, at which party how much CPU time is consumed. For comparability, each party uses the same CPU and is limited to one core. 
Figures~\ref{img:t1calc} and \ref{img:t2calc} confirm that centralized and decentralized calculations do not burden the smartphone of the customer. With such transactions, the customer is responsible for generating evaluation keys and decrypting values, which are fast operations (Figure~\ref{img:operations}).

Figure~\ref{img:t1calc} corresponds to a centralized transaction where any computation and relinearization takes place at the mobility platform. Encrypting values must be performed at the data sources. With the mobility providers as data sources, this corresponds to an edge computing scenario, where all time-consuming operations are executed on a centralized or decentralized cloud instance. 

Figure~\ref{img:t2calc} transfers any computationally expensive operation in the domain of the mobility providers. If we take a m6i.32xlarge cloud instance for comparison, each transaction costs each mobility provider less than 0,003~millicents, even with the slowest FHE scheme. 
Figure~\ref{img:t3calc} confirms that 
even customer-side calculations incur negligible overhead on the customer's smartphone. 
Similarly to T1 and T2, with T3 the encryption of the data takes place at the mobility providers. 

Note that for to calculations with unencrypted values at the cost of privacy, the runtimes and the costs of a single calculation are below measurement accuracy, and virtually zero. Thus, FHE is not suitable for any big-data problem, or for scenarios where numerous transactions must be executed to the smallest possible costs. However, in the field of smart mobility, such transaction fees are several orders of magnitude smaller than the billing amount on the customer's invoice, but privacy is an important factor. Thus, we have confirmed that FHE schemes are feasible for  business models in the field of smart mobility,

\paragraph{Approximated Memory Comsumption}
In order to have a practical estimate of the memory consumption incurred by FHE, we have implemented each customer, mobility platform and mobility provider as an individual application. Thus, we have measured the total amount of memory of the application, libraries, runtime variables and the buffers where encrypted values are stored. The isolated increase in buffer sizes needed to store encrypted intermediate results can be found in \cite{cryptoeprint:2011/277}(BFV), \cite{cryptoeprint:2012/144}(BGV) and \cite{cheon2017homomorphic}(CKKS). Table~\ref{tab:memconsmpt} summarizes this.

With our experiments, we measured a memory consumption between 207~MB and 321~MB for BFV. We measured between 216~MB and 306~MB for BGV, and 146~MB to 188~MB for CKKS. A large memory consumption corresponds to more expensive operations (encryption and calculations including relinearization). This was because such operations require many runtime variables and, therefore, a large and deep stack. In comparison, an execution on unencrypted values resulted in applications with a memory footprint between 76~MB and 104~MB. Thus, none of the FHE schemes utilized memory resources that exceeded even the capacity of a smartphone.\\[0.5cm]

\begin{table}[ht]
	\centering
		\begin{tabular}[h]{|c|c|}\hline
            \textsl{FHE Scheme} & \textsl{Memory Consumption} \\\hline
Plain Text & 76 -- 104 MB \\\hline
BFV & 207 -- 321 MB \\\hline
BGV & 216 -- 306 MB \\\hline
CKKS & 146 -- 188 MB \\\hline
		\end{tabular}
  \vspace{15pt}
    \caption{\label{tab:memconsmpt} Approximated Memory Consumption}
\end{table}


\section{Conclusion}
\label{sec:conclusion}

In the upcoming years, the planning of cities and transportation logistics for moving people and goods will undergo significant changes. The conventional concept of mobility using individual transportation modes, such as a car, is not longer useful due to environmental reasons and growing cities. The demand for multi-modal transport solutions that allow users to move flexible and eco-friendly is high. However, the implementation of this approach requires the sharing of sensitive personal data with various parties, creating potential privacy risks. 

This paper explored the potential use of fully homomorphic encryption as an efficient and noise-free solution for data privacy concerns in the implementation of smart mobility. Initially, privacy requirements for such a smart mobility approach were formulated, based on which three multiparty computations were identified that benefit from FHE. An implementation was provided using state-of-the-art FHE schemes BGV~\cite{cryptoeprint:2012/144}, BFV~\cite{cryptoeprint:2011/277} and CKKS~\cite{cheon2017homomorphic} based on Microsoft SEAL~\cite{sealcrypto23}. Finally, memory consumption and execution times were measured, evaluated and compared with a non-encrypted benchmark. To provide optimal experimental results, a benchmark framework was used to monitor memory consumption and execution times. To test the applicability of FHE in a real-life smart mobility scenario, the ressources used in the implementation were analyzed and compared to available resources on smartphones and cloud instances.  

Based on the experiments conducted, encrypting transactions with FHE increases CPU time by approximately 100~milliseconds compared to unencrypted transactions. However, this additional processing time does not adversely affect the user experience~\cite{nielsen2013mobile}. The use of parallel processing can significantly reduce this time, and the cost of such encryption on a current cloud instance is less than 3 microcents. We conclude that FHE is a cost-effective means of ensuring privacy, and a viable option for a smart mobility business model.

For future research, it would be beneficial to scale the implementation to a real-life scenario involving multiple smartphones functioning as edge devices, and leveraging cloud instances for both the mobility platform and service providers. Hence, experiments could be extended to measure actual runtimes including side effects of operating systems, and delays of a virtualization environment. Also, delays of network connection could be reported. Furthermore, it would be worthwhile to evaluate and compare other libraries such as OpenFHE \cite{al2022openfhe} and other state-of-the-art FHE schemes.

To conclude, this paper demonstrated the feasibility and practicality of FHE in smart mobility scenarios and emphasized its potential as a solution for privacy concerns associated with the sharing of sensitive personal data.




\bibliographystyle{IEEEtran}
\bibliography{papers}

\begin{thebibliography}{10}
\providecommand{\url}[1]{#1}
\csname url@samestyle\endcsname
\providecommand{\newblock}{\relax}
\providecommand{\bibinfo}[2]{#2}
\providecommand{\BIBentrySTDinterwordspacing}{\spaceskip=0pt\relax}
\providecommand{\BIBentryALTinterwordstretchfactor}{4}
\providecommand{\BIBentryALTinterwordspacing}{\spaceskip=\fontdimen2\font plus
\BIBentryALTinterwordstretchfactor\fontdimen3\font minus
  \fontdimen4\font\relax}
\providecommand{\BIBforeignlanguage}[2]{{%
\expandafter\ifx\csname l@#1\endcsname\relax
\typeout{** WARNING: IEEEtran.bst: No hyphenation pattern has been}%
\typeout{** loaded for the language `#1'. Using the pattern for}%
\typeout{** the default language instead.}%
\else
\language=\csname l@#1\endcsname
\fi
#2}}
\providecommand{\BIBdecl}{\relax}
\BIBdecl

\bibitem{jevinger2019potentials}
{\AA}.~Jevinger and J.~A. Persson, ``Potentials of context-aware travel support
  during unplanned public transport disturbances,'' \emph{Sustainability},
  vol.~11, no.~6, p. 1649, 2019.

\bibitem{al2021enabling}
A.~Al-Rahamneh \emph{et~al.}, ``Enabling customizable services for multimodal
  smart mobility with city-platforms,'' \emph{IEEE Access}, vol.~9, pp.
  41\,628--41\,646, 2021.

\bibitem{schuppan2014urban}
J.~Schuppan, S.~Kettner, A.~Delatte, and O.~Schwedes, ``Urban multimodal travel
  behaviour: Towards mobility without a private car,'' \emph{Transportation
  Research Procedia}, vol.~4, pp. 553--556, 2014.

\bibitem{chowdhury2018preference}
M.~S. Chowdhury, M.~A. Osman, and M.~M. Rahman, ``Preference-aware public
  transport matching,'' in \emph{2018 International Conference on Innovation in
  Engineering and Technology (ICIET)}.\hskip 1em plus 0.5em minus 0.4em\relax
  IEEE, 2018, pp. 1--6.

\bibitem{broring2017enabling}
A.~Br{\"o}ring, S.~Schmid, C.-K. Schindhelm, A.~Khelil, S.~K{\"a}bisch,
  D.~Kramer, D.~Le~Phuoc, J.~Mitic, D.~Anicic, and E.~Teniente, ``Enabling iot
  ecosystems through platform interoperability,'' \emph{IEEE software},
  vol.~34, no.~1, pp. 54--61, 2017.

\bibitem{li2017pare}
Y.~Li \emph{et~al.}, ``Pare: A system for personalized route guidance,'' in
  \emph{Conference on World Wide Web}, 2017.

\bibitem{herzog2017routeme}
D.~Herzog, H.~Massoud, and W.~W{\"o}rndl, ``Routeme: A mobile recommender
  system for personalized, multi-modal route planning,'' in \emph{Conference on
  User Modeling, Adaptation and Personalization}, 2017.

\bibitem{campigotto2016personalized}
P.~Campigotto, C.~Rudloff, M.~Leodolter, and D.~Bauer, ``Personalized and
  situation-aware multimodal route recommendations: the favour algorithm,''
  \emph{IEEE Transactions on Intelligent Transportation Systems}, vol.~18,
  no.~1, pp. 92--102, 2016.

\bibitem{moran2017hybrid}
O.~Moran, R.~Gilmore, R.~Ord{\'o}{\~n}ez-Hurtado, and R.~Shorten, ``Hybrid
  urban navigation for smart cities,'' in \emph{2017 IEEE 20th International
  Conference on Intelligent Transportation Systems (ITSC)}.\hskip 1em plus
  0.5em minus 0.4em\relax IEEE, 2017, pp. 1--6.

\bibitem{paiva2020privacy}
S.~Paiva \emph{et~al.}, ``Privacy and security challenges in smart and
  sustainable mobility,'' \emph{SN Applied Sciences}, vol.~2, pp. 1--10, 2020.

\bibitem{borchersprivacy}
T.~Borchers \emph{et~al.}, ``Privacy concerns on the mobility of smart
  cities,'' in \emph{Brazilian Technology Symposium (BTSym’21)}, 2021.

\bibitem{de2022impact}
E.~P. de~Mattos \emph{et~al.}, ``The impact of mobility on location privacy,''
  \emph{IEEE Systems Journal}, vol.~16, no.~4, pp. 5509--5520, 2022.

\bibitem{eckhoff2017privacy}
D.~Eckhoff and I.~Wagner, ``Privacy in the smart city,'' \emph{IEEE
  Communications Surveys \& Tutorials}, vol.~20, no.~1, pp. 489--516, 2017.

\bibitem{zhao2020survey}
P.~Zhao, G.~Zhang, S.~Wan, G.~Liu, and T.~Umer, ``A survey of local
  differential privacy for securing internet of vehicles,'' \emph{The Journal
  of Supercomputing}, vol.~76, pp. 8391--8412, 2020.

\bibitem{khaliq2022secure}
A.~A. Khaliq, A.~Anjum, A.~B. Ajmal, J.~L. Webber, A.~Mehbodniya, and S.~Khan,
  ``A secure and privacy preserved parking recommender system using elliptic
  curve cryptography and local differential privacy,'' \emph{IEEE Access},
  vol.~10, pp. 56\,410--56\,426, 2022.

\bibitem{qin2022toward}
G.~Qin, S.~Deng, Q.~Luo, J.~Sun, and H.~Kerivin, ``Toward privacy-aware
  multimodal transportation: Convergence to network equilibrium under
  differential privacy,'' \emph{Available at SSRN 4244002}, 2022.

\bibitem{shanthi2014efficient}
P.~Shanthi and S.~Balasundaram, ``An efficient clique cloak algorithm for
  defending location-dependent attacks in location based services,'' in
  \emph{Conference on Information and Communication Technology for Competitive
  Strategies}, 2014.

\bibitem{memon2018pseudonym}
I.~Memon, L.~Chen, Q.~A. Arain, H.~Memon, and G.~Chen, ``Pseudonym changing
  strategy with multiple mix zones for trajectory privacy protection in road
  networks,'' \emph{International Journal of Communication Systems}, vol.~31,
  no.~1, p. e3437, 2018.

\bibitem{martelli2022price}
F.~Martelli, M.~E. Renda, and J.~Zhao, ``The price of privacy control in
  mobility sharing,'' in \emph{Sustainable Smart City Transitions}.\hskip 1em
  plus 0.5em minus 0.4em\relax Routledge, 2022, pp. 233--258.

\bibitem{li2019autompc}
T.~Li, L.~Lin, and S.~Gong, ``Autompc: Efficient multi-party computation for
  secure and privacy-preserving cooperative control of connected autonomous
  vehicles.'' in \emph{SafeAI@ AAAI}, 2019.

\bibitem{raja2020ai}
G.~Raja \emph{et~al.}, ``Ai-powered blockchain-a decentralized secure
  multiparty computation protocol for iov,'' in \emph{IEEE Conference on
  Computer Communications}, 2020.

\bibitem{acar2018survey}
A.~Acar, H.~Aksu, A.~S. Uluagac, and M.~Conti, ``A survey on homomorphic
  encryption schemes: Theory and implementation,'' \emph{ACM Computing
  Surveys}, vol.~51, no.~4, pp. 1--35, 2018.

\bibitem{sealcrypto23}
{Microsoft Research, Redmond, WA.}, ``{M}icrosoft {SEAL} (release 4.1),''
  \url{https://github.com/Microsoft/SEAL}, Jan. 2023, accessed Feb. 20th, 2023.

\bibitem{helib23}
S.~Halevi \emph{et~al.}, ``{HElib 2.2.2, December 2022},''
  \url{https://github.com/homenc/HElib}, 2023, accessed Feb. 20th, 2023.

\bibitem{OpenFHE23}
{OpenFHE.}, ``{OpenFHE},'' \url{https://www.openfhe.org/}, 2023, accessed Feb.
  20th, 2023.

\bibitem{cryptoeprint:2012/144}
\BIBentryALTinterwordspacing
J.~Fan and F.~Vercauteren, ``Somewhat practical fully homomorphic encryption,''
  Cryptology ePrint Archive, Paper 2012/144, 2012. [Online]. Available:
  \url{https://eprint.iacr.org/2012/144}
\BIBentrySTDinterwordspacing

\bibitem{cryptoeprint:2011/277}
\BIBentryALTinterwordspacing
Z.~Brakerski, C.~Gentry, and V.~Vaikuntanathan, ``Fully homomorphic encryption
  without bootstrapping,'' Cryptology ePrint Archive, Paper 2011/277, 2011.
  [Online]. Available: \url{https://eprint.iacr.org/2011/277}
\BIBentrySTDinterwordspacing

\bibitem{cheon2017homomorphic}
J.~H. Cheon, A.~Kim, M.~Kim, and Y.~Song, ``Homomorphic encryption for
  arithmetic of approximate numbers,'' in \emph{Conference on the Theory and
  Applications of Cryptology and Information Security}, 2017.

\bibitem{nielsen2013mobile}
J.~Nielsen and R.~Budiu, \emph{Mobile usability}.\hskip 1em plus 0.5em minus
  0.4em\relax MITP-Verlags GmbH \& Co. KG, 2013.

\bibitem{gentry2009fully}
C.~Gentry, \emph{A fully homomorphic encryption scheme}.\hskip 1em plus 0.5em
  minus 0.4em\relax Stanford university, 2009.

\bibitem{regev2006lattice}
O.~Regev, ``Lattice-based cryptography,'' in \emph{Advances in
  Cryptology-CRYPTO 2006: 26th Annual International Cryptology Conference,
  Santa Barbara, California, USA, August 20-24, 2006. Proceedings 26}.\hskip
  1em plus 0.5em minus 0.4em\relax Springer, 2006, pp. 131--141.

\bibitem{lyubashevsky2013ideal}
V.~Lyubashevsky, C.~Peikert, and O.~Regev, ``On ideal lattices and learning
  with errors over rings,'' \emph{Journal of the ACM (JACM)}, vol.~60, no.~6,
  pp. 1--35, 2013.

\bibitem{behera2022design}
S.~Behera and J.~R. Prathuri, ``Design of novel hardware architecture for fully
  homomorphic encryption algorithms in fpga for real-time data in cloud
  computing,'' \emph{IEEE Access}, vol.~10, pp. 131\,406--131\,418, 2022.

\bibitem{gupta2022memfhe}
S.~Gupta \emph{et~al.}, ``Memfhe: End-to-end computing with fully homomorphic
  encryption in memory,'' \emph{ACM Transactions on Embedded Computing
  Systems}, 2022.

\bibitem{lee2022privacy}
J.-W. Lee, H.~Kang, Y.~Lee, W.~Choi, J.~Eom, M.~Deryabin, E.~Lee, J.~Lee,
  D.~Yoo, Y.-S. Kim \emph{et~al.}, ``Privacy-preserving machine learning with
  fully homomorphic encryption for deep neural network,'' \emph{IEEE Access},
  vol.~10, pp. 30\,039--30\,054, 2022.

\bibitem{chen2021privacy}
J.~Chen, K.~Li, and S.~Y. Philip, ``Privacy-preserving deep learning model for
  decentralized vanets using fully homomorphic encryption and blockchain,''
  \emph{IEEE Transactions on Intelligent Transportation Systems}, vol.~23,
  no.~8, pp. 11\,633--11\,642, 2021.

\bibitem{wibawa2022homomorphic}
F.~Wibawa \emph{et~al.}, ``Homomorphic encryption and federated learning based
  privacy-preserving cnn training: Covid-19 detection use-case,'' in
  \emph{European Interdisciplinary Cybersecurity Conference}, 2022.

\bibitem{stripelis2021secure}
D.~Stripelis \emph{et~al.}, ``Secure neuroimaging analysis using federated
  learning with homomorphic encryption,'' in \emph{Symposium on Medical
  Information Processing and Analysis}, vol. 12088, 2021, pp. 351--359.

\bibitem{zhu2019privacy}
L.~Zhu, M.~Li, Z.~Zhang, C.~Xu, R.~Zhang, X.~Du, and N.~Guizani,
  ``Privacy-preserving authentication and data aggregation for fog-based smart
  grid,'' \emph{IEEE Communications Magazine}, vol.~57, no.~6, pp. 80--85,
  2019.

\bibitem{ren2021privacy}
W.~Ren \emph{et~al.}, ``Privacy-preserving using homomorphic encryption in
  mobile iot systems,'' \emph{Computer Communications}, vol. 165, pp. 105--111,
  2021.

\bibitem{baharon2015new}
M.~R. Baharon \emph{et~al.}, ``A new lightweight homomorphic encryption scheme
  for mobile cloud computing,'' in \emph{IEEE Computer and Information
  Technology}, 2015.

\bibitem{gdpr}
{Council of the European Union}, ``Regulation (eu) 2016/679 on the protection
  of natural persons with regard to the processing of personal data and on the
  free movement of such data,'' OJ L 119, 4.5.2016, p. 1–88, 2016.

\bibitem{fairphone23}
\BIBentryALTinterwordspacing
{Fairphone B.V}, ``{Fairphone 4},'' 2022, accessed Feb. 20th, 2023. [Online].
  Available: \url{https://www.fairphone.com/}
\BIBentrySTDinterwordspacing

\bibitem{amazonaws23}
\BIBentryALTinterwordspacing
{Amazon Web Services, Inc.}, ``{Amazon EC2 M6i Instances},'' 2023, accessed
  Feb. 20th, 2023. [Online]. Available:
  \url{https://aws.amazon.com/de/ec2/instance-types/}
\BIBentrySTDinterwordspacing

\bibitem{doctest22}
\BIBentryALTinterwordspacing
V.~Kirilov \emph{et~al.}, ``Doctest v2.4.9,'' 2022, accessed Feb. 20th, 2023.
  [Online]. Available: \url{https://github.com/doctest/doctest}
\BIBentrySTDinterwordspacing

\bibitem{log4cplus22}
\BIBentryALTinterwordspacing
T.~E. Smith \emph{et~al.}, ``log4cplus v2.1.0,'' 2023, accessed Feb. 20th,
  2023. [Online]. Available: \url{https://github.com/log4cplus}
\BIBentrySTDinterwordspacing

\bibitem{googlebench22}
\BIBentryALTinterwordspacing
{Google Inc.}, ``google/benchmark v1.7.1,'' 2022, accessed Feb. 20th, 2023.
  [Online]. Available: \url{https://github.com/google/benchmark}
\BIBentrySTDinterwordspacing

\bibitem{vtune}
{Intel Corporation}, ``{Intel VTune Profiler},'' \url{https://www.intel.com},
  2023, accessed Mar. 07th, 2023.

\bibitem{al2022openfhe}
A.~Badawi \emph{et~al.}, ``Openfhe: Open-source fully homomorphic encryption
  library,'' in \emph{Encrypted Computing \& Applied Homomorphic Cryptography},
  2022.

\end{thebibliography}


\end{document}